# Pseudomagnetic Control of Light Waves in the Electrically Tunable Photonic Crystals with Deformation Engineering

*Zhipeng Qi, Hao Sun, Guohua Hu\*, Xiumin Song, Yaohui Sun, Wanghua Zhu, Bo Liu\*, Xuechao Yu, Francois M. Peeters\*, and Yiping Cui\**

Dr. Z. Qi, Dr. X. Song, Prof. B. Liu, Prof. F. M. Peeters

School of Physics and Optoelectronic Engineering, Nanjing University of Information Science & Technology, Nanjing 210044, China

E-mail: bo@nuist.edu.cn; francois.peeters@uantwerpen.be

Dr. H. Sun

The Institute for Functional Intelligent Materials (I-FIM), National University of Singapore, Singapore 117544, Singapore

Prof. G. Hu, Y. Sun, W. Zhu, Prof. Y. Cui

Advanced Photonics Center, School of Electronic Science and Engineering, Southeast University, Nanjing 210096, China

E-mail: photonics@seu.edu.cn; cyp@seu.edu.cn

Prof. B. Liu

Department of Electrical Engineering, Tsinghua University, Beijing 100876, China

Prof. X. Yu

Key Laboratory of Multifunctional Nanomaterials and Smart Systems, Suzhou Institute of Nano-Tech and Nano-Bionics, Chinese Academy of Sciences, Suzhou 215123, China

Prof. F. M. Peeters

Department of Physics, University of Antwerp, Groenenborgerlaan 171, B-2020 Antwerp, Belgium

Funding: This work is supported by the National Natural Science Foundation of China (62225503, 62105158).

**Abstract**: With the demonstrations of pseudo-magnetism in optical systems, the pursuits of its practical applications require not only the use of pseudomagnetic fields to create functional optical devices but also a reliable method to manipulate pseudo-magnetism-affected light waves. Here, we experimentally demonstrate an ultracompact Si-based cavity formed by triaxially deformed photonic honeycomb lattices. The triaxial deformation could lead to Landau quantization, showing the possibilities of realizing the localization and resonating of photons with pseudomagnetic fields. Through adopting the Si waveguides for directional coupling, we successfully obtain the transmission spectra for the proposed cavities in the photonic integrated circuits. This opens a novel avenue for highly efficient excitations and detections of Landau-quantized photonic density of states totally on chip. Moreover, we verify a linear electrical tunability of -0.018 THz/mW for the pseudo-magnetism-induced optical resonant states, fulfilling the manipulation of photons without varying deformations. Our work introduces a mechanism for performing tunable light waves in triaxial deformation-engineered systems, which enriches the design principles of integrated optical devices.

Z. Qi and H. Sun contributed equally to this work.

# 1. Introduction

Pseudomagnetic fields (PMFs) give rise to unique physical phenomena in the two-dimensional (2D) electron systems, providing an alternative route to investigate magnetism-related effects. For example, PMFs function as real magnetic fields, enabling observations of quantum hall effect (QHE) without breaking time-reversal symmetry[1]. In the case of PMFs acting on Dirac Fermions, Landau quantization can be easily achieved, contributing to discrete energy states as well as flat bands. More intriguingly, the strength of PMF could reach up to several hundred Tesla, which is orders of magnitude larger than the strength of real magnetic field[2–6]. With the creation of strong PMFs, it is available to modify the physical properties of 2D materials, including the electron mobility[7,8], valley scattering[9,10], carrier relaxation[11], and second-order susceptibility[12], to a naturally unattainable extent.

Essentially, PMFs originate from the non-uniform synthetic gauge fields: $\boldsymbol{B}_p = \nabla \times \boldsymbol{A}_s$. In addition to electrons, the synthetic gauge field $\boldsymbol{A}_s$ also offers opportunities in the manipulation of neutral particles[13,14], which is promising for the implementation of topological phases in solid-state systems based on classical waves[15,16]. Recently, the periodic modulations on structures, known as Floquet engineering[17], have been proposed to yield $\boldsymbol{A}_s$, successfully exerting the magnetic-like effects on both photons[18–23] and phonons[24–27]. In the past decades, graphene has been playing a crucial role in modern physics due to its gapless and relativistic band structure[28], providing an ideal platform to investigate $\boldsymbol{A}_s$[29]. As predicted by Kane and Mele[30], $\boldsymbol{A}_s$ could be produced by applying a certain mechanical strain onto the graphene flake. To mimic the strain engineering in graphene, huge amounts of effort are devoted to the study of lattice systems with aperiodic deformations[31–37], prompting the experimental verifications on PMFs for light or sound waves[38–46]. Recent studies have developed a class of unconventional optical resonators inspired by the photonic simulations of exotic electronic states (e.g., defect states and valley kink states), including L3[47] and valley[48] photonic crystal (PhC) cavities. These devices exhibit significantly improved performances, such as small modal volumes, high quality ($Q$) factors, and disorder robustness. Although the PMF has shown great promise for realizing one-way propagative optical waveguides[42,49], there remains a lack of experimental demonstrations regarding the localization and resonating of photons in PMF-based optical systems, not to mention the tunability of PMF-induced optical resonances, which limits its potential applications in functional integrated optical devices.

Here, we propose a novel on-chip optical resonator based on triaxially deformed PhCs for the pseudomagnetic control of light waves. In our study, the photonic honeycomb lattices whose Dirac points are located within the telecom band are patterned with triaxial deformations on the silicon-on-insulator (SOI) substrate. The PMF, induced by the interactions between deformed lattices, leads to Landau-quantized energy states with closed orbits for photons. By exploiting such PhCs to build photonic cavities, we unveil the mechanism to confine the light waves with the PMF that requires neither real magnetic fields nor conventional optical reflections. With the integrations of Si waveguides for directional coupling, it is convenient to obtain the optical responses of such deformation-engineered PhC cavities, which are related to the PMF-modified density of states (DOS), totally on chip.

To demonstrate the tunability of light waves in the triaxially deformed optical systems, we fabricate metallic electrodes onto the top layer of the proposed cavity, where the PMF-induced optical resonances are sensitive to the refractive indices (*RI*s) of the surrounding areas. Without altering the PMF via structural perturbations[36], we demonstrate the electrical modulations of optical resonances based on thermo-optic (TO) effect under a fixed PMF. In comparison with other tuning methods, the TO approach features easy fabrication, low optical loss, low cost, high complementary metal-oxide semiconductor (CMOS) compatibility, and high integration degree. Particularly, the high-efficiency TO modulation is more readily achievable in ultracompact PhC cavities, making it an ideal choice for our study. This work not only enriches graphene physics in the optical domain but connects PMFs to electrically tunable PhCs, providing a conceptual route to achieve active photonic integrated circuits (PICs) based on pseudo-magnetisms[36].

## 2. Results and Discussion
### 2.1. Applying triaxial deformations on photonic honeycomb lattices

As depicted in **Figure 1**a, we start from the large-scale photonic graphene with $C_{6v}$ symmetry[50] consisting of circular $SiO_2$ holes in a Si slab. The entire fabrication process is compatible with the CMOS technology. The dispersion for the in-plane transverse electric (TE) waves features a Dirac degeneracy at the *K* or *K′* point (see Supplementary Section S1 for details). Around the Dirac point of 195.3 THz, the low-energy photonic dispersion meets: $\omega(\mathbf{k}_{K/K'} \pm \delta\mathbf{k}) = \omega_{K/K'} \pm v_D|\delta\mathbf{k}|$, where $\omega_{K/K'}$ is the angular frequency for the *K* or *K′* point and $v_D$ is the group velocity.

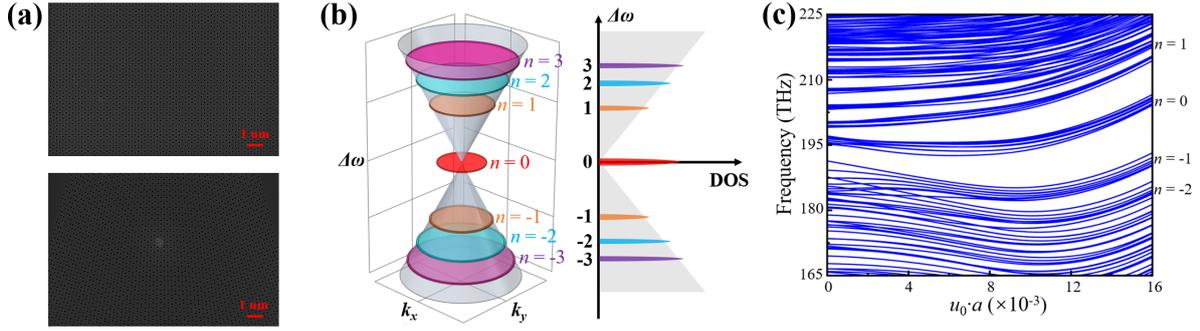

**Figure 1**. PMF-induced Landau quantization. (a) Scanning electron micrographs (SEMs) of the pristine (top) and triaxially deformed (bottom) photonic graphene on an SOI substrate. (b) Schematic illustration showing Landau quantization under the PMFs. The Dirac cone is split into quantized Landau levels (left), with the continuous photonic DOS rearranged into discrete energy states (right). (c) Numerically obtained eigenfrequencies as a function of $u_0 \cdot a$, labelled with photonic Landau levels (PLLs) of $n$ = -2, -1, 0, and 1. $a$ is the lattice constant, which is a fixed value of 408 nm. The calculations were carried out for SOI-based photonic graphene composed of 729 lattices.

To apply an artificial strain onto the pristine photonic graphene, a lithographic mask is designed in a way that each sublattice is arranged to a new location based on an in-plane position-dependent displacement $\boldsymbol{u}(\boldsymbol{r}) = [u_x(x, y), u_y(x, y)]$, which varies with the position vector $\boldsymbol{r}$. Such lattice deformation leads to a spatial-dependent coupling coefficient of photon $t_{ij}(\boldsymbol{r}_i, \boldsymbol{r}_j)$[51], where $\boldsymbol{r}_i$ and $\boldsymbol{r}_j$ are the positions of the $A$ and $B$ sublattices, respectively. Generally, $t_{ij}$ depends on the nearest-sublattice distance $\rho_{ij}$, which offers an opportunity to induce $\boldsymbol{A}_s(\boldsymbol{r}) = [A_x(x, y), A_y(x, y)]$ with the modifications of $\boldsymbol{u}(\boldsymbol{r})$[31,36,38,40]. A uniform PMF can be achieved when the lattices are displaced from their original locations as follows: $u_x(x, y) = 2u_0 xy$ and $u_y(x, y) = u_0(x^2-y^2)$, where $u_0$ denotes the deformation value (see Supplementary Section S3 for details). This contributes to the varying shifts of Dirac cones in both momentum and frequency[52]. And the resulting PMF is given by

$$\boldsymbol{B}_p = \left\{ \partial_x A_y(x, y) - \partial_y A_x(x, y) \right\} \cdot \hat{z} = \xi \frac{4\beta u_0}{a_0} \cdot \hat{z}, \tag{1}$$

where $\hat{z}$ is the unit vector normal to the plane. $|\boldsymbol{B}_p|$ is proportional to $u_0$ with a constant $\beta = |\partial \log t / \partial \log a_0| \approx 2$ and displays opposite signs for the $K$ ($\xi = +1$) and $K'$ ($\xi = -1$) valleys because of conserved time-reversal symmetry[29]. Influenced by the PMF, the relativistic band structure is rearranged into non-equidistant discrete energy states (see Figure 1b), of which angular frequencies close to the Dirac point follow:

$$\omega_n = \omega_{K/K'} + \text{sgn}(n) v_D \sqrt{\frac{|n\boldsymbol{B}_p|}{2}}, \qquad (2)$$

where $n$ is Landau order, which is an integer. In our system, the photonic DOS could reveal the total intensity of PMF-affected photonic states summed over all lattice sites[32]. In Figure 1c, it is found that the triaxial deformation could induce the reorganization of energy states for the SOI-based photonic graphene, with the observation of $0^{th}$ PLL even under a small $u_0$ of $0.004/a$. Once $u_0$ is increased above $0.008/a$, we can see the well-defined PLLs of $n$ = -2, -1, and 0, showing distinct energy gaps that are devoid of discrete states. At $u_0 = 0.016/a$, the PLL of $n$ =1 can be roughly identified, which bends away from the Dirac point.

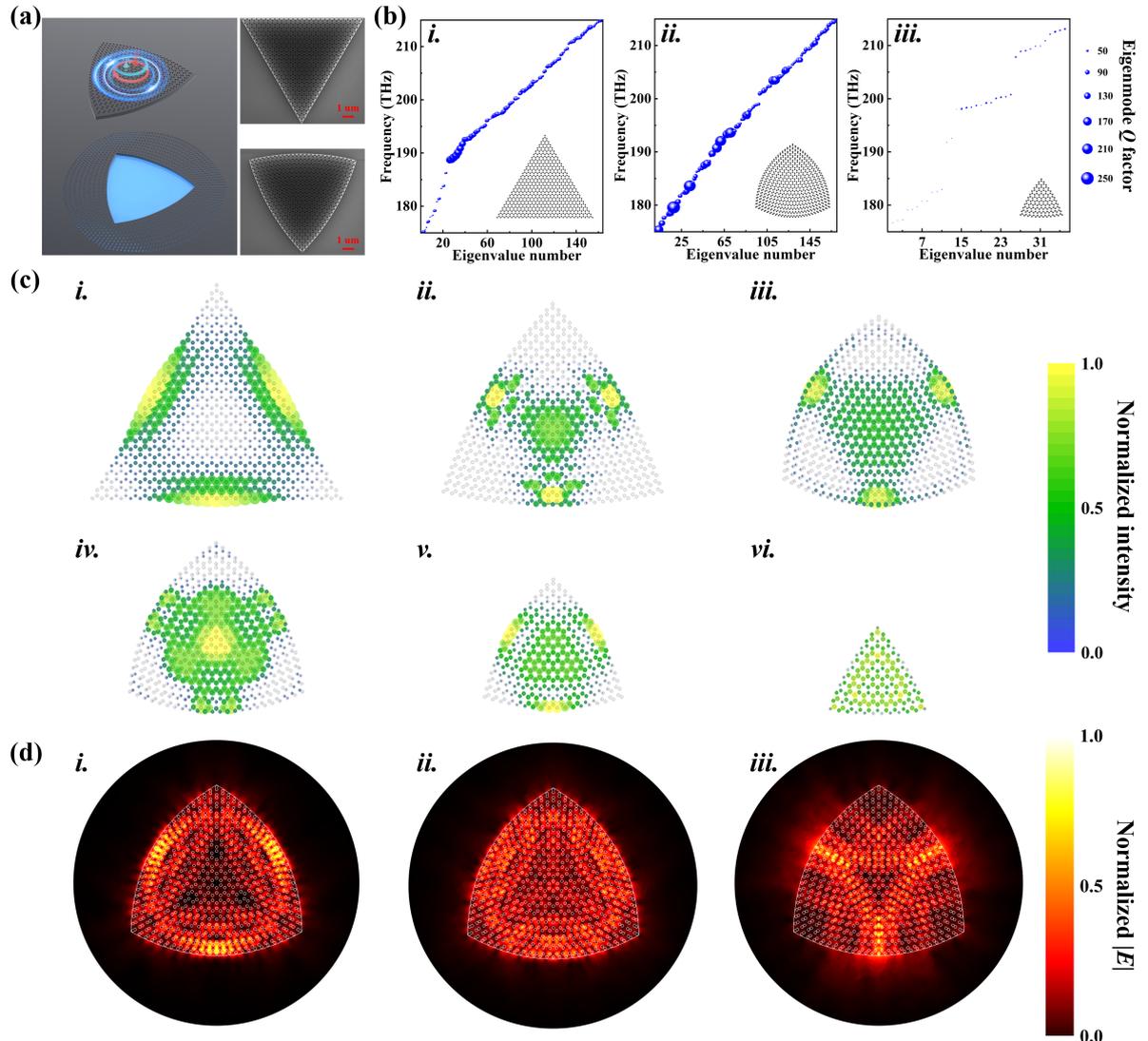

**Figure 2**. PhC cavities engineered by the triaxial deformations. (a) Left panel: Schematic of the PhC cavity truncated from the triaxially deformed photonic graphene along the zigzag edges, where the light waves can be localized by the PMF that distributes uniformly, and is perpendicular to the surface. Right panel: SEMs of the undeformed (top) and triaxially deformed (bottom) PhC cavities, with *m* denoting the number of hexagonal unit cells arranged along the shortest distance from the center to the edge of the cavity. (b) Numerically computed eigenvalues with *Q* factors indicated by the size of blue ball for the structural parameters of *i*. $m = 4.5$, $u_0 = 0$; *ii*. $m = 4.5$, $u_0 = 0.016/a$; *iii*. $m = 2$, $u_0 = 0.016/a$, respectively. Insets: Schematics of the photonic honeycomb lattices with different $u_0$. (c) Numerically calculated zero-energy wavefunction intensities for the proposed cavities with: (*i - iii*) $m = 4.5$ and varying $u_0$ of *i*. 0, *ii*. $0.008/a$ and *iii*. $0.016/a$; (*iv - vi*) $u_0 = 0.016/a$ and varying *m* of *iv*. 4, *v*. 3, and *vi*. 2. Both color and area of the circles correspond to the intensity on the photonic lattices. (d) Normalized electric field distributions of optical eigenmodes in the cavity of $m = 4.5$ and $u_0 = 0.016/a$, corresponding to *i*. $f = 190.72$ THz, $Q = 223$; *ii*. $f = 191.61$ THz, $Q = 135$; *iii*. $f = 192.53$ THz, $Q = 42$.

As shown in **Figure 2**a, photons in the triaxially deformed PhC cavity could be deflected into the PMF-induced closed orbits terminated with the zigzag edges. This is not only suitable for the ultracompact integrations but accessible for the experimental detections of PMF-affected photonic DOS[36,40]. Figure 2b shows the calculated eigenvalues for the cavities with different deformations (*i.e.*, $u_0$) and sizes (*i.e.*, *m*), in ascending order, within the frequency range of 175 ~ 215 THz, revealing that the PMF-affected photonic states are unable to constitute flat bands because of the hybridizations with zigzag edge states. As confirmed by the studies on triaxially strained microwave graphene[40], these states would gradually reorganize into a sequence of Landau levels with the increase of $u_0$. In Figure 2c, one can observe the intensity profiles of the wavefunctions corresponded to the Dirac point and $0^{th}$ PLLs (see Figure 2c (*i - iii*)) change greatly in that they move from the edge into the bulk regions when $u_0$ is increased from 0 to $0.016/a$, showing that the increased PMF could boost mode localization. Simultaneously, the triaxial deformation engineering could slightly reduce the footprint of the PhC cavity. Compared to the other dielectric cavities based on PhCs[53] or optical waveguides[54], our approach enables both the enhancement of mode localization and the miniaturization of devices without material (e.g., high refractive index) and fabrication (e.g., smooth sidewalls) refinements.

The hybridizations of bulk and edge states could be strengthened with the reduction in cavity size (see Figure 2c (*iv - vi*)), causing the degradation in mode localization. As *m* decreases from 4 to 2, there are few distinctions between the edge and bulk states since they are highly hybridized, contributing to a near-uniform spatial distribution of zero-energy wavefunction intensity throughout the cavity with $u_0$ set to be 0.016/*a*. Besides the bulk states, the zigzag edge states are also subject to the deformation-induced PMFs. As seen in Figure 2d, the zigzag edges for the cavity of *m* = 4.5 can be tailored by the large triaxial deformation of $u_0$ = 0.016/*a* to generate closed orbits for photons, greatly suppressing the radiative losses of edge states. This gives rise to a high-*Q* optical eigenmode at 190.7 THz with its field mainly localized within the edge part of the cavity, which is analogous to the optical resonant mode in the micro-ring resonator. As a result of the enhancements in intercoupling between edge and bulk states, especially at 192.53 THz, PMF-induced optical eigenmodes suffer from weakened localization, intensifying their radiative leakages into the surrounding area, and therefore lowering *Q* factors.

## 2.2. Localization and resonating of photons under the PMFs

In this work, we take advantage of PICs based on the integrations of triaxial deformed PhC cavities and Si waveguides to facilitate the experiments (see **Figure 3**a). The PMF-induced optical resonant states can be excited by the TE waves in the Si waveguide (see Supplementary Figure S3). As shown in Figure 3b, a number of honeycomb lattices are fabricated to correspond to increasing the deformation values, from $u_0$ = 0 to 0.016/*a* in a step length of 0.004/*a*. For $u_0$ = 0, the photonic states are a continuum with a conical Dirac dispersion near the *K* and *K'* points, which can be attributed to the relativistic nature of the graphene-like band structure. In Figure 3c, we find that the peak weights of PMF-induced resonant states are enhanced when $u_0$ increases, along with the appearance of more peaks in the DOS spectra. The number of photons localized in the cavity can be reflected in a measure of the transmitted optical power, which is related to the eigenmode damping rate $\eta$ and the coupling coefficient $\kappa$ between the cavity and the Si waveguide. A high transmitted power corresponds to the delocalized case (*i.e.*, low DOS), whereas a low one indicates that the input TE waves are strongly localized by the cavity with few energies transmitted through, corresponding to the high DOS. Additionally, the linewidth of the transmission spectral curve is decided by the total *Q* factor of such Si waveguide-coupled cavity, which comprises the eigenmode *Q* factor ($\propto 1/\eta$) as well as the coupling *Q* factor ($\propto 1/\kappa$)[54]. The total *Q* factor is inversely proportional to the total decay rate ($\kappa+\eta$).

As shown in Figure 3d, the spectrum measured for the undeformed PhC cavity ($u_0 = 0$) is featureless without any resonant dips, which can be attributed to the absence of PMF. Meanwhile, the spectrum recorded over the cavity with $u_0 = 0.004/a$ displays a subtly modified curve compared to the undeformed one, where, however, no clear signatures of optical resonances can be observed in the detection window. As $u_0$ is increased above $0.008/a$, the transmission spectra exhibit a succession of resonant dips with exotic features, including inequivalent spacings and amplitudes. For the resonances based on the coherent nature of travelling light waves, the observed dips would be expected to follow a different feature, most notably showing periodic distributions in the spectra with fixed free spectral ranges similar to the whispering gallery modes. And our PhC cavities did not show any evidence of defects or disorders. Consequently, these resonant dips in the transmission spectra are induced by the deformation-induced PMFs. This also indicates that the PMF-induced optical resonant states within the cavities can be directly excited by the TE waves in Si waveguides, which correlate with the photonic DOS.

$\kappa$ exhibits strong frequency dependence, arising from the variations in spatial localization of different PMF-induced optical eigenmodes. For the PhC cavity of $u_0 = 0.016/a$ and $m = 4.5$, we find that both $\kappa$ and $\eta$ are very small and close to each other near the DOS peak frequency of 193.5 THz, which mainly benefits from the strong mode localization. In contrast, $\kappa$ increases greatly at off-DOS peak frequencies due to the degraded mode localization (*i.e.*, enhanced mode radiations). This further reduces the contribution weights of these off-peak photonic states under the waveguide coupling conditions because of their large $\kappa$, hence yielding a transmission dip narrower than the DOS peak. Figure S8 depicts a series of simulated transmission spectra for different $u_0$, which are in good agreement with the measured ones. Also, the coupling between Si waveguide and PhC cavity has been numerically studied. In addition to an extreme low coupling loss (~ 1 dB), a high transmission contrast (> 35 dB) can be obtained for such waveguide-coupled cavity (see Supplementary Figure S5). In comparison to the waveguide coupling with L3[55] or valley[56] PhC cavities, our coupling efficiency is much higher and even comparable to those of bent coupler-based Si micro-rings[57], which is highly suitable for Si-based photonic integrations.

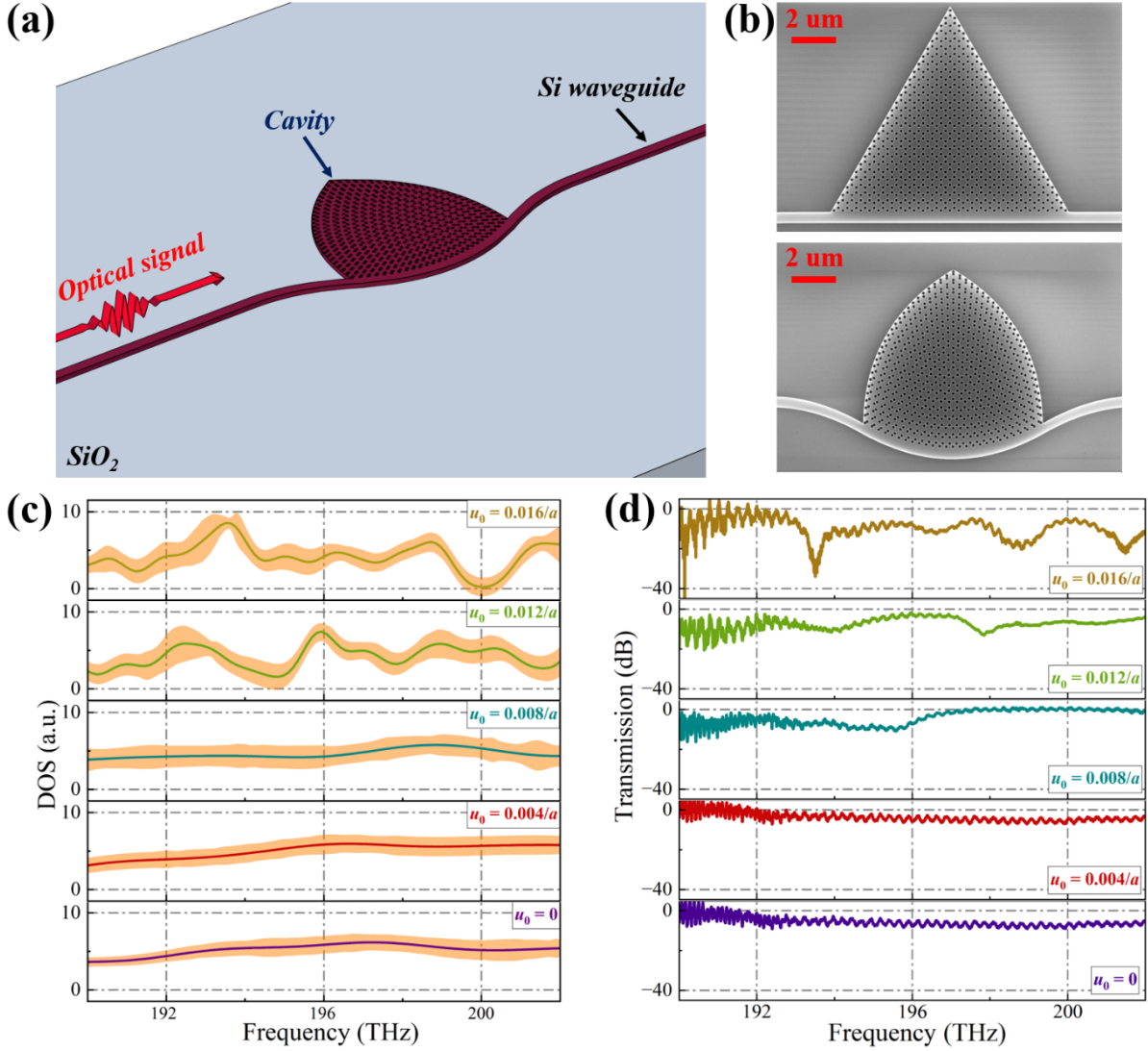

**Figure 3.** Triaxially deformed PhC cavity coupled with the Si waveguide. (a) Schematic of a deformation-engineered PhC cavity with the integration of a Si waveguide, where the PMF-induced optical resonant states can be excited by the TE waves through directional coupling. (b) SEMs of the PhC cavities, which are directional-coupled with Si waveguides along their zigzag edges for the same $m = 4.5$ but different $u_0$ of 0 (top) and $0.016/a$ (bottom), respectively. (c) Calculated photonic DOS with $\kappa$ indicated by the orange shading and (d) measured transmission spectra of the proposed cavities for $u_0$ setting to be 0, $0.004/a$, $0.008/a$, $0.012/a$, and $0.016/a$, respectively. With the increase of $u_0$, more photonic DOS peaks and transmission dips emerge in the spectra.

In the quantum Hall systems, Landau waves could only spread out over a certain distance[58]. With the consideration of semiclassical limits, the radius of PMF-induced cyclotron orbit $R_c = v_D^2 m_c/|F_p|$, with the effective cyclotron mass $m_c = \Delta\omega/v_D^2$ and the Lorentz-like force $\boldsymbol{F}_p = \xi \boldsymbol{v_D} \times \boldsymbol{B}_p$. As illustrated by Figure 2b (*ii.*) and (*iii.*), the total number of eigenstates is mainly determined by the cavity size, which can be estimated by the shortest distance from its center to edge: $L = 3ma_0 \cdot (1 + 3ma_0u_0)$. And we can observe that the eigenmode $Q$ factors decrease with the reduction in cavity size, even though the cavities own the same $u_0$, which is caused by the degradation in localization of photons. To systematically investigate the influences of cavity size on the spectral characteristics (e.g., formations and variations) of PMF-induced optical resonances, we fabricate a series of PhC cavities with the same deformation ($u_0 = 0.004/a$) but different sizes ($m$ varying from 5 to 7.5 in a step of 0.5) on an SOI platform (see **Figure 4**a). One can observe the transmission spectra ordered with increasing $m$ underline the resonant properties of PMF-governed light waves (see Figure 4b). Once $m$ is increased above 5.5, a transmission dip can be observed in the spectrum, manifesting that the cavity size is large enough to support the cyclotron motions of photons for the case of $u_0 = 0.004/a$. According to the standing wave equation, the angular cut-off frequency shift of the optical cavity should meet: $\Delta\omega = l \cdot v_D / (2L)$, and $l = 0, 1, 2\cdots\cdots$. Here, the linear relationship between the angular frequency shift of the PMF-induced optical resonant state and the reciprocal of the cavity size, as depicted in Figure 4c, evidences the direct proportionality between $\Delta\omega$ and $1/L$. And the increase of $L$ could not only help to prevent mode leakages but also provide more available PMF orbits so as to localize more photons, which accounts for the increase in extinction ratio.

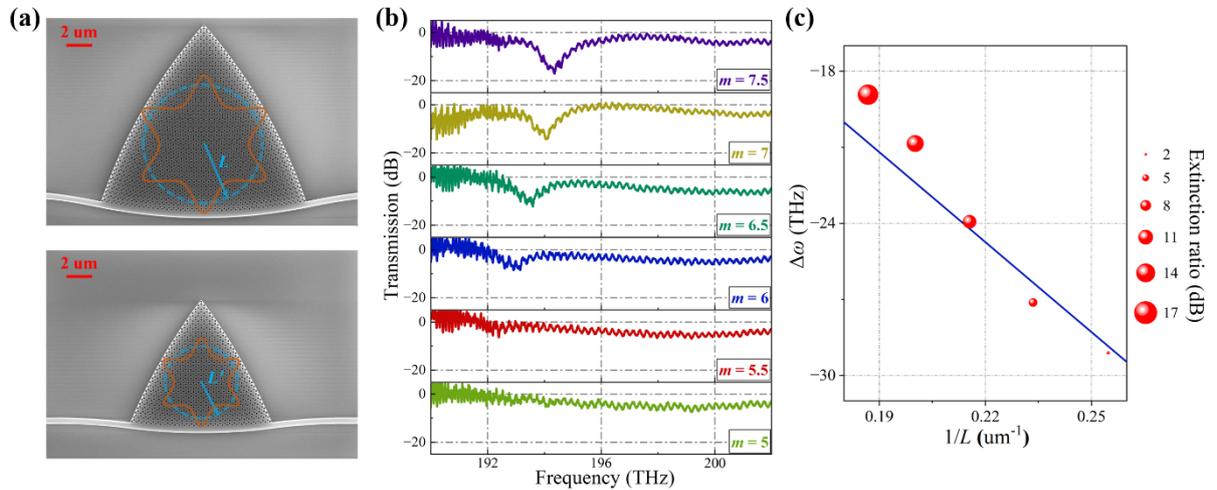

**Figure 4**. Size effects on PMF-induced optical resonances. (a) SEMs of the waveguide-coupled PhC cavities, for the same $u_0 = 0.004/a$ and different $m$ of 7.5 (top) and 5 (bottom). The closed cyclotron orbits for the localized photons are highlighted in blue, with the radii decided by $L$. The optical resonant states are highlighted in yellow, which exist in the form of standing waves. (b) Measured transmission spectra of the PhC cavities with $m$ set to be 5, 5.5, 6, 6.5, 7, and 7.5, respectively. They are fabricated with the same $u_0$ of $0.004/a$. (c) The frequency shift relative to the Dirac point of the PMF-induced optical resonances versus the reciprocal of $L$. Red dots:

Experimentally extracted frequency shifts of the transmission dips shown in (b). Cyan solid line: Theoretically calculated $\Delta\omega$ is a linear function of $1/L$.

## 2.3. Electrical manipulations of PMF-governed light waves

To verify the dynamic control of light waves in the triaxially deformed PhC cavities, we go beyond directly altering the structural parameters, which dominate for the magnitudes of PMFs and cyclotron orbits of photons within the cavity. Instead, we define an effective refractive index factor $N_{eff}$, which is highly sensitive to the surrounding areas. Generally, $v_D = c / N_{eff}$, where $c$ denotes the speed of light in the vacuum. Thus, the photonic states for the proposed cavities can be modulated through varying $N_{eff}$. Here, we can obtain

$$\frac{\partial \omega_n}{\partial P} = -\frac{c}{N_{eff}^2} \cdot \frac{\partial T}{\partial P} \cdot \frac{\partial N_{eff}}{\partial T} \left( |\mathbf{k}_{K/K'}| + \mathrm{sgn}(n)\sqrt{\frac{|n\mathbf{B}_p|}{2}} \right), \tag{3}$$

where $P$ is the electrical (or heating) power and $T$ is the temperature of PhC cavity. Each Landau quantized angular frequency $\omega_n$ is proportional to $P$ under the condition that the variation in $N_{eff}$ is much smaller than itself. In **Figure 5**a, we show the microscopic images of the electrically tunable PhC cavity of $m = 4.5$ and $u_0 = 0.016/a$ with the integration of a Si waveguide. Once a certain voltage is applied to the electrodes, the $RI$s of Si and $SiO_2$ layers could be varied by the Joule heat because of the TO effect, affecting the optical responses.

To verify the electrical tunability of PMF-governed light waves, we establish an experimental setup as schematically shown in Figure S4. We utilize a direct current (DC) power supply to apply voltages ($V_{DC}$) onto the metallic electrode pads with two probes. Figure S6 exhibits the simulated distributions of heat in the proposed configuration with different $P$. Compared to the temperature variation ($\Delta T$) in $SiO_2$ upper cladding layer, it is found that $\Delta T$ distributes uniformly in the proposed cavity, which plays a dominant role in the manipulation of $N_{eff}$. This is because the TO coefficient of Si is one order of magnitude larger than that of $SiO_2$ (see Methods). And the TE-polarized optical eigenmodes are mostly confined in the PhC with only a few energies spreading into the upper and lower cladding layers. As depicted in Figure S7, both $\Delta T$ and $N_{eff}$ are proportional to $P$, revealing that the PMF-affected photonic states can be precisely manipulated by varying $P$.

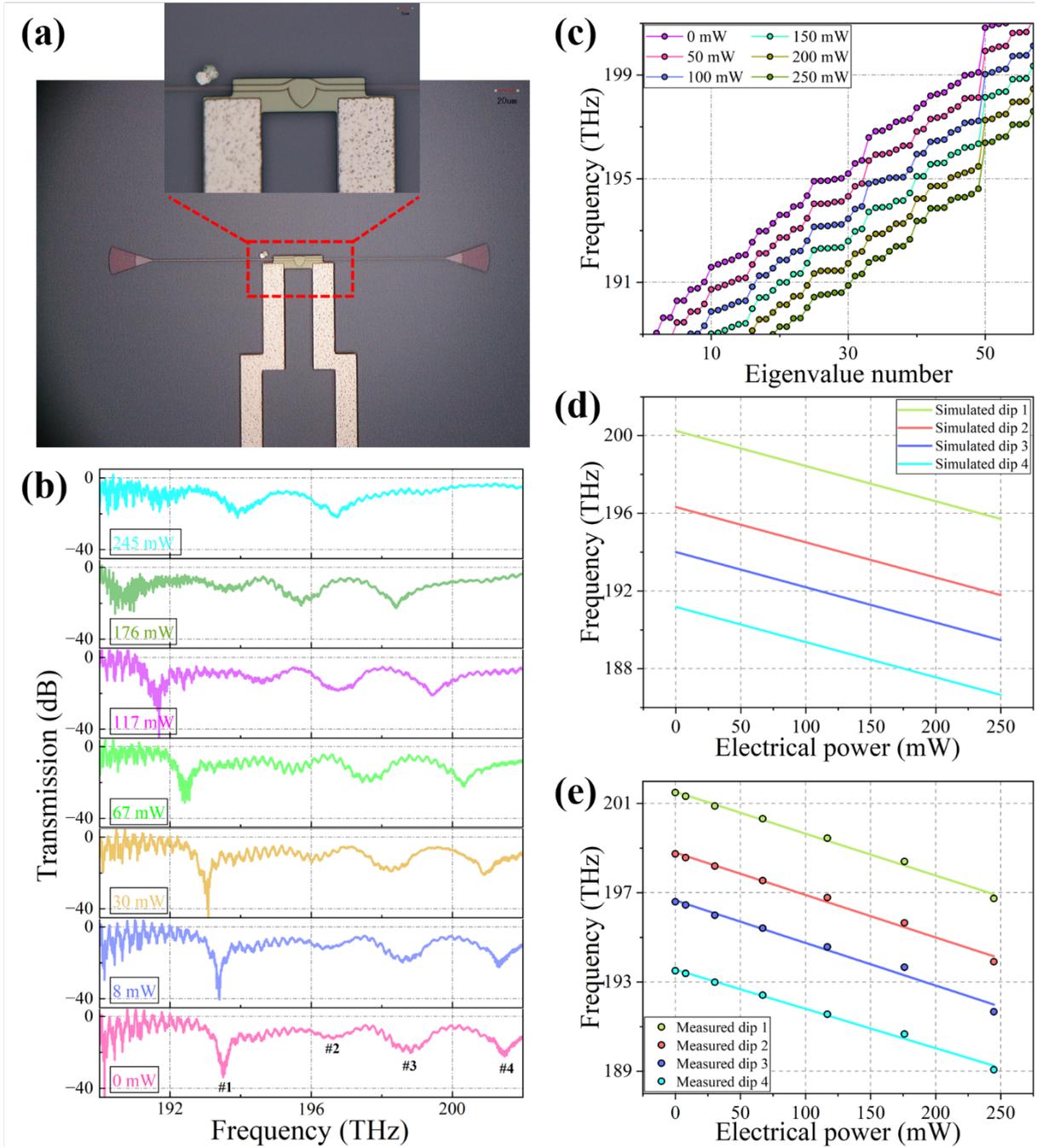

**Figure 5.** Electrical tunability of PMF-governed light waves. (a) Microscope images of the electrically tunable integrated optical device composed of a PhC cavity with $m = 4.5$ and $u_0 = 0.016/a$, a bent Si waveguide, two Si grating couplers, and a metallic electrode. An enlarged image shows that the heating electrode is fabricated on the top of $SiO_2$ upper cladding layer, of which dimensions are large enough to cover the cavity. (b) Measured transmission spectra with the electrical power increasing from 0 to 245 mW. The resistance of the entire electrical connection is around 22 Ω. (c) Numerically computed eigenvalues are plotted in ascending order with the increase of electrical power from 0 to 250 mW in a step of 50 mW. Both current-induced Joule heat and TO effect are considered in the model for eigenvalue simulations. (d)

Theoretically and (e) experimentally extracted optical resonant frequencies in the transmission spectra near the $0^{th}$ PLL versus the electrical power. The measured results in (e) are indicated by solid dots and fitted with a linear scaling law (solid lines).

For the case of $u_0 = 0.016/a$ and $m = 4.5$, there is a sequence of PMF-induced resonant dips emerging in the transmission spectra within the Dirac region. As shown in Figure 5b, there are negligible variations in resonance dip spacings during TO modulations because the PMF is fixed. Figure 5c presents the calculated eigenvalues with different $P$, showing that PMF-affected photonic states can be effectively modulated by an electrical approach instead of altering the structural parameters of the cavity. As $P$ is increased, there is an obvious redshift for the experimental transmission spectra, which matches well with the simulated results (see Figures 5d and e). Moreover, the electrical tunability of such cavity could reach nearly -0.018 THz/mW, which ranks relatively high among the integrated TO devices due to its small size and ultracompact structure. This modulation efficiency could be further improved by decreasing upper cladding layer thickness and introducing air trenches or substrate undercuts[59]. The successful TO modulation in the triaxially deformed PhC cavity paves the way for the applications of other tuning methods, such as carrier injection/depletion and optical pumping, onto the analogous photonic devices. It would be very promising to utilize deformation engineering to fabricate active PMF-based optical devices on an SOI platform, featuring flexible tunability, ultracompact size, and unique resonant properties.

## 3. Conclusion

In summary, we have experimentally demonstrated pseudomagnetic control of light waves in the PhC cavities via deformation engineering. By mimicking the triaxially strained graphene, we successfully induce PMFs in the photonic honeycomb lattices for the localization and resonating of photons on an SOI substrate, which offers discrete photonic states with Landau quantized orbits. This study enriches the physics of optical resonant systems, in which the spectral responses arise from the PMFs. The PMF-induced Landau quantization provides possibilities for enhancing light emissions and nonlinear optic effects, which is highly welcomed for the on-chip lasing[31], as the confined optical modes corresponding to the discrete energy states could be selectively enhanced by employing site-dependent damping[34]. Besides, the integrations of deformation-engineered PhC cavities and Si waveguides make it possible to realize the on-chip interconnections and electrical modulations based on PMF-modified photonic DOS, which is attractive for the development of novel integrated optical devices. This work can also be extended to the other classic wave systems, facilitating the practical applications of PMFs in acoustics and polaritons[39,41].

**4. Methods**

*Simulations*: 3D finite-element method (FEM) simulations were conducted by using the COMSOL Multiphysics software. The *RI*s of Si and $SiO_2$ were set to be 3.446 and 1.446, respectively. The core layer was Si slab, with a thickness of 220 nm. The upper and lower cladding layers were $SiO_2$, with thicknesses of 2 and 1.8 um. The undeformed rhombic unit cell consisted of two circular $SiO_2$ holes with a diameter of 136 nm and a lattice constant of 408 nm. Perfectly matched layers (PMLs) were set as the boundaries that surround the simulation domain, enabling the estimates for the frequencies, *Q* factors, and spatial distributions of the optical eigenmodes. Here, the *Q* factor was defined as the ratio between the real part and twice the imaginary part of the complex eigenfrequency. TE waves were considered for the simulations.

Besides, we also perform multi-physics simulations in which the heat field is introduced into the 3D model, which could in turn vary the *RI* of each layer. The thermal insulation was used as the boundary condition for the interfaces of the simulation domain far away from the PhC cavity, where no heat flux across the boundaries. Without applied voltages, the device operates at room temperature, which was set to be 293.15 K. The TiN stripe acts as the heat source, whose temperature can be modulated by supplying a certain electrical power to it. During the heating, the bottom Si substrate was regarded as the heat sink. There was heat flux $q_0 = h \cdot (T_{ext} - T)$ across the boundaries between the upper cladding layer and the air, where *h* denotes the heat transfer coefficient, $T_{ext}$ is the external temperature, and *T* is the temperature of the upper cladding layer. In this model, $h = 10$ W/(m$^2$/K) and $T_{ext} = 293.15$ K. The *RI*s of Si and $SiO_2$ could be directly influenced by the current-induced $\Delta T$, which are highly related to the heat distributions in the device configuration. The TO coefficients of Si and $SiO_2$ were set as $1.8 \times 10^{-4}$ /K and $1 \times 10^{-5}$ /K, respectively. More details are provided in Supplementary section S5.

Full-wave simulations on the transmissions for the proposed PICs were conducted by using Lumerical FDTD solutions software. The PhC cavities were modeled in as same way as the ones used in the Comsol simulations. The Si waveguide has a rectangular cross-section, with a width of 500 nm and a height of 220 nm. PMLs were set as the boundaries surrounding the devices. The mode source was used to launch the TE-polarized light waves ranging from 180

~ 202 THz into the Si waveguide on one side, and a monitor was set to detect the transmitted power on the other side of the waveguide.

*Sample fabrication*: The SOI substrate was patterned using electron beam lithography (EBL) and a dry plasma etching approach. Patterns for the Si-based honeycomb lattices, waveguides, and grating couplers were first defined in a 350 nm-thick polymethyl methacrylate (PMMA) layer, which was prepared by spin coating. With PMMA acting as an etching mask, all the patterns were transferred to the 220 nm-thick top Si layer by inductively coupled plasma (ICP) etching. Each rhombic unit cell is comprised of two cylindrical holes with a nearest-neighbor spacing $a_0$ of 235 nm. The Si waveguide has dimensions of $500 \times 220$ nm$^2$ (width × height). And the Si grating is designed to be a focusing structure with the total length of 39.26 um, the section angle of 36.2°, the first grating line span of 10.04 um, the fiber tilt angle of 85°, the period of 0.655 um, and the fill factor of 0.15. Then the whole chip was coated with a 1.6 um-thick SiO$_2$ upper cladding layer by plasma-enhanced chemical vapor deposition (PECVD).

For the purpose of electrical modulations, the metallic electrodes were fabricated by lift-off process. Through taking advantage of photomask for alignment, patterns for the heat part of electrodes were defined in a 1.8 um-thick spin-coated photoresist layer based on UV lithography. Next, the 100 nm-thick TiN layer was evaporated onto the chip. After removal of the residual photoresist layer with acetone, the TiN strip of $45 \times 12$ um$^2$ (length × width) was fabricated on top of the PhC cavity. And the patterned Al layers were prepared in the same way, which was used to connect the TiN part with the electrode pad. The electrode pad was made by a 100 nm-thick Cu layer, of which dimensions are $180 \times 80$ um$^2$ (length × width). Finally, a 200 nm-thick SiO$_2$ layer was deposited onto the chip so as to bury the electrodes for protection. In order to apply voltages onto the electrode pads, the $176 \times 76$ um$^2$ (length × width) windows were opened on the top of pad regions by UV photolithography and ICP etching. Further details have been provided in Supplementary section S2.

*Experimental setup*: Optical measurements of the Si waveguide-coupled PhC cavities were conducted by using the fiber-chip alignment system. A tunable *c.w.* laser (TSL-710, Santec) was utilized to launch the light waves into the single-mode fiber (SMF), sweeping from 1480 nm to 1640 nm with a wavelength resolution of 0.01 nm during the measurement. Before coupling into the PICs, the fiber mode was modulated by a polarization controller (PC) so that only the fundamental TE mode could be excited in the Si waveguide. Moreover, two electrical probes with micro-scale tips are used to contact the electrode pads on the chip, which could introduce the tunable currents with a DC voltage source (E36312A, Keysight). Finally, the transmitted light waves can be captured by the output SMF and then detected by a power meter (MPM-210, Santec). The detailed measurement setup has been illustrated in Supplementary section S4.

## Supporting Information

Supporting Information is available from the Wiley Online Library or from the author.

## Acknowledgements

Z. Qi and H. Sun contributed equally to this work. This work is supported by the National Natural Science Foundation of China (62225503, 62105158).

## Data Availability Statement

All data are available from the corresponding author upon reasonable request.

## References

[1] F. Guinea, M. I. Katsnelson, A. K. Geim, *Nat. Phys.* **2010**, *6*, 30.

[2] N. Levy, S. A. Burke, K. L. Meaker, M. Panlasigui, A. Zettl, F. Guinea, A. H. C. Neto, M. F. Crommie, *Science* **2010**, *329*, 544.

[3] M. Neek-Amal, L. Covaci, F. M. Peeters, *Phys. Rev. B* **2012**, *86*, 041405.

[4] S. Zhu, J. A. Stroscio, T. Li, *Phys. Rev. Lett.* **2015**, *115*, 245501.

[5] P. Jia, W. Chen, J. Qiao, M. Zhang, X. Zheng, Z. Xue, R. Liang, C. Tian, L. He, Z. Di, X. Wang, *Nat. Commun.* **2019**, *10*, 3127.

[6] C. C. Hsu, M. L. Teague, J. Q. Wang, N. C. Yeh, *Sci. Adv.* **2020**, *6*, eaat9488.

[7] N. J. G. Couto, D. Costanzo, S. Engels, D. K. Ki, K. Watanabe, T. Taniguchi, C. Stampfer, F. Guinea, A. F. Morpurgo, *Phys. Rev. X* **2014**, *4*, 1.

[8] L. Wang, P. Makk, S. Zihlmann, A. Baumgartner, D. I. Indolese, K. Watanabe, T. Taniguchi, C. Schönenberger, *Phys. Rev. Lett.* **2020**, *124*, 157701.

[9] D. R. Da Costa, A. Chaves, G. A. Farias, L. Covaci, F. M. Peeters, *Phys. Rev. B* **2012**, *86*, 115434.

[10] P. Kun, G. Kukucska, G. Dobrik, J. Koltai, J. Kürti, L. P. Biró, L. Tapasztó, P. Nemes-Incze, *npj 2D Mater. Appl.* **2019**, *3*, 11.

[11] D. H. Kang, H. Sun, M. Luo, K. Lu, M. Chen, Y. Kim, Y. Jung, X. Gao, S. J. Parluhutan, J. Ge, S. W. Koh, D. Giovanni, T. C. Sum, Q. J. Wang, H. Li, D. Nam, *Nat. Commun.* **2021**, *12*, 5087.

[12] K. Lu, M. Luo, W. Gao, Q. J. Wang, H. Sun, D. Nam, *Nat. Commun.* **2023**, *14*, 2580.

[13] M. Hafezi, *Int. J. Mod. Phys. B* **2014**, *28*, 7.

[14] M. Aidelsburger, S. Nascimbene, N. Goldman, *Comptes Rendus Phys.* **2018**, *19*, 394.

[15] L. Lu, J. D. Joannopoulos, M. Soljačić, *Nat. Photonics* **2014**, *8*, 821.

[16] Z. Yang, F. Gao, X. Shi, X. Lin, Z. Gao, Y. Chong, B. Zhang, *Phys. Rev. Lett.* **2015**, *114*, 114301.

[17] M. Bukov, L. D'Alessio, A. Polkovnikov, *Adv. Phys.* **2015**, *64*, 139.

[18] M. Hafezi, E. A. Demler, M. D. Lukin, J. M. Taylor, *Nat. Phys.* **2011**, *7*, 907.

[19] R. O. UmucalIlar, I. Carusotto, *Phys. Rev. A* **2011**, *84*, 043804.

[20] K. Fang, Z. Yu, S. Fan, *Nat. Photonics* **2012**, *6*, 782.

[21] M. C. Rechtsman, J. M. Zeuner, Y. Plotnik, Y. Lumer, D. Podolsky, F. Dreisow, S. Nolte, M. Segev, A. Szameit, *Nature* **2013**, *496*, 196.

[22] S. Mittal, J. Fan, S. Faez, A. Migdall, J. M. Taylor, M. Hafezi, *Phys. Rev. Lett.* **2014**, *113*, 087403.

[23] L. J. Maczewsky, J. M. Zeuner, S. Nolte, A. Szameit, *Nat. Commun.* **2017**, *8*, 13756.

[24] N. Swinteck, S. Matsuo, K. Runge, J. O. Vasseur, P. Lucas, P. A. Deymier, *J. Appl. Phys.* **2015**, *118*, 063103.

[25] R. Fleury, A. B. Khanikaev, A. Alù, *Nat. Commun.* **2016**, *7*, 11744.

[26] J. P. Mathew, J. del Pino, E. Verhagen, *Nat. Nanotechnol.* **2020**, *15*, 198.

[27] Z. Cheng, R. W. Bomantara, H. Xue, W. Zhu, J. Gong, B. Zhang, *Phys. Rev. Lett.* **2022**, *129*, 254301.

[28] A. H. Castro Neto, F. Guinea, N. M. R. Peres, K. S. Novoselov, A. K. Geim, *Rev. Mod. Phys.* **2009**, *81*, 109.

[29] M. A. H. Vozmediano, M. I. Katsnelson, F. Guinea, *Phys. Rep.* **2010**, *496*, 109.

[30] C. L. Kane, E. J. Mele, *Phys. Rev. Lett.* **1997**, *78*, 1932.

[31] H. Schomerus, N. Y. Halpern, *Phys. Rev. Lett.* **2013**, *110*, 013903.

[32] G. Salerno, T. Ozawa, H. M. Price, I. Carusotto, *2D Mater.* **2015**, *2*, 34015.

[33] Z. Yang, F. Gao, Y. Yang, B. Zhang, *Phys. Rev. Lett.* **2017**, *118*, 194301.

[34] H. Abbaszadeh, A. Souslov, J. Paulose, H. Schomerus, V. Vitelli, *Phys. Rev. Lett.* **2017**, *119*, 195502.

[35] C. Brendel, V. Peano, O. J. Painter, F. Marquardt, *Proc. Natl. Acad. Sci. U. S. A.* **2017**, *114*, E3390.

[36] C. R. Mann, S. A. R. Horsley, E. Mariani, *Nat. Photonics* **2020**, *14*, 669.

[37] Z. T. Huang, K. Bin Hong, R. K. Lee, L. Pilozzi, C. Conti, J. S. Wu, T. C. Lu, *Nanophotonics* **2022**, *11*, 1297.

[38] M. C. Rechtsman, J. M. Zeuner, A. Tünnermann, S. Nolte, M. Segev, A. Szameit, *Nat. Photonics* **2013**, *7*, 153.

[39] X. Wen, C. Qiu, Y. Qi, L. Ye, M. Ke, F. Zhang, Z. Liu, *Nat. Phys.* **2019**, *15*, 352.

[40] M. Bellec, C. Poli, U. Kuhl, F. Mortessagne, H. Schomerus, *Light Sci. Appl.* **2020**, *9*, 146.

[41] O. Jamadi, E. Rozas, G. Salerno, M. Milićević, T. Ozawa, I. Sagnes, A. Lemaître, L. Le Gratiet, A. Harouri, I. Carusotto, J. Bloch, A. Amo, *Light Sci. Appl.* **2020**, *9*, 144.

[42] Z. Qi, H. Sun, G. Hu, C. Deng, W. Zhu, B. Liu, Y. Li, S. Liu, X. Yu, Y. Cui, *Photonics Res.* **2023**, *11*, 1294.

[43] R. Barczyk, L. Kuipers, E. Verhagen, *Nat. Photonics* **2024**, *18*, 574.

[44] M. Barsukova, F. Grisé, Z. Zhang, S. Vaidya, J. Guglielmon, M. I. Weinstein, L. He, B. Zhen, R. McEntaffer, M. C. Rechtsman, *Nat. Photonics* **2024**, *18*, 580.

[45] B. Yang, X. Shen, L. Shi, Y. Yang, Z. H. Hang, *Adv. Photonics Nexus* **2024**, *3*, 026011.

[46] J. Guglielmon, M. C. Rechtsman, M. I. Weinstein, *Phys. Rev. A* **2021**, *103*, 013505.

[47] Y. Akahane, T. Asano, B. S. Song, S. Noda, *Nature* **2003**, *425*, 944.

[48] J. W. Dong, X. D. Chen, H. Zhu, Y. Wang, X. Zhang, *Nat. Mater.* **2017**, *16*, 298.

[49] H. Jia, M. Wang, S. Ma, R. Y. Zhang, J. Hu, D. Wang, C. T. Chan, *Light Sci. Appl.* **2023**, *12*, 165.

[50] O. Peleg, G. Bartal, B. Freedman, O. Manela, M. Segev, D. N. Christodoulides, *Phys. Rev. Lett.* **2007**, *98*, 103901.

[51] Y. Chen, X. Chen, X. Ren, M. Gong, G. C. Guo, *Phys. Rev. A* **2021**, *104*, 023501.

[52] M. Neek-Amal, L. Covaci, K. Shakouri, F. M. Peeters, *Phys. Rev. B* **2013**, *88*, 115428.

[53] R. M. De La Rue, C. Seassal, *Laser Photonics Rev.* **2012**, *6*, 564.

[54] S. Feng, T. Lei, H. Chen, H. Cai, X. Luo, A. W. Poon, *Laser Photonics Rev.* **2012**, *6*, 145.

[55] R. Bose, D. Sridharan, G. S. Solomon, E. Waks, *Opt. Express* **2011**, *19*, 4204.

[56] Y. F. Gao, Y. H. He, Y. Li, S. Rouzi, M. C. Jin, Y. He, S. Y. Zhou, *Opt. Laser Technol.* **2024**, *175*, 110799.

[57] P. Chen, S. Chen, X. Guan, Y. Shi, D. Dai, *Opt. Lett.* **2014**, *39*, 6304.

[58] M. Jamotte, N. Goldman, M. Di Liberto, *Commun. Phys.* **2022**, *5*, 1.

[59] M. Kim, Y. Ban, F. Ferraro, D. Coenen, N. Rajasekaran, P. Verheyen, R. Magdziak, H. Sar, P. De Heyn, D. Velenis, P. Absil, J. Van Campenhout, *J. Light. Technol.* **2025**, *43*, 1328.

This work presents a new way to control light waves on the SOI substrate with the adoptions of deformation-engineered photonic crystals, enabling the resonating of photons due to the pseudomagnetic fields. Through coupling with Si waveguides and leveraging thermo-optic effect, electrically tunable pseudo-magnetism-induced optical resonant states can be achieved in the photonic integrated circuits, supplying a new scheme for optoelectronic integrations.

**Pseudomagnetic Control of Light Waves in the Electrically Tunable Photonic Crystals with Deformation Engineering**

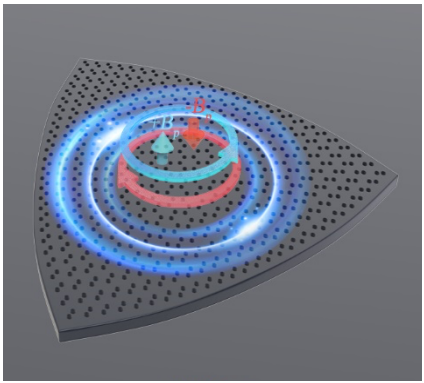